\documentstyle[12pt]{article}
\begin{document}
%%%%%%%%%%%%

\begin{titlepage}
\begin{flushright}
       {\bf UK/93-04}\\
        August 1993 \\
      hep-lat/9308015
\end{flushright}
\begin{center}

{\bf {\LARGE  Stochastic Estimation with $Z_2$ Noise}}

\vspace{1.5cm}

{\bf    Shao-Jing Dong and Keh-Fei Liu }\\ [0.5em]
{\it  Dept. of Physics and Astronomy  \\
  Univ. of Kentucky, Lexington, KY 40506}
 \\[1em]

\end{center}

\vspace{1cm}

\begin{abstract}
We introduce a $Z_2$ noise for the stochastic estimation of matrix
inversion and discuss its superiority over other noises including the
Gaussian noise. This algorithm is applied to the calculation of
quark loops in lattice quantum chromodynamics that involves
diagonal and off-diagonal traces of the inverse matrix. We will
point out its usefulness in its applications to
estimating determinants, eigenvalues, and eigenvectors, as well as
its limitations based on the structure of the inverse matrix.

\vspace{0.8cm}
\noindent

\end{abstract}

\vfill
\end{titlepage}

% and here comes the text ...

Either formally or as a result of numerical practicality,
many physical systems, be they classical or quantum mechanical,
are boiled down to solving matrix equations. As the dimension
$N \times N$ of the matrix M approaches the limit of
solving the matrix equation $MX = S$ computationally
for a source S with dimension
$N \times 1$, it becomes unfeasible to solve it for an S of
dimension $N \times N$ using the same algorithm. This is so simply
because it requires N times more computational
time as that of solving for a column of S. In physics applications,
sometimes one needs to compute quantities which amounts to solving
for the whole matrix S, e.g. calculating the diagonal and off-diagonal
traces of $M^{-1}$. This poses a numerical challenge,
sometimes a grand one. Especially hard is when the dimension of the
matrix M grows fast with the physical variables of the problem. For
example, when M is represented in the space-time coordinates,
the dimension N grows as $L^4$ where L is the size of the space-time
dimension. Quark matrix in lattice
guage calculation of quantum chromodynamics (QCD) falls in this
category. A space-time
lattice with merely the size of $16^3 \times 24$ gives a quark
matrix of the dimension $10^6 \times 10^6$ including the
spin and color degrees of freedom. While it is durable to calculate
the quark propagator, i.e. $M^{-1}(x,0)$ for a point source S at 0
with a reasonably small quark mass (e.g. a fraction of the strange
quark mass) on today's supercomputers, the quark propagator
$M^{-1}(x,y)$ from any point to any point is certainly unattainable.
For calculations of the 2-point functions or
3-point functions with direct
insertions, one can get by with the help of translational
symmetry. But there are cases where one can not rely on such a help.
These include the calculations of quark loops which
are space-time or space integrations of the fermion propagators.
Examples of interest in QCD include the quark condensate and the
topological susceptibility with the fermion method~\cite{sv87},
flavor-singlet
meson masses which involve disconnected quark loops in the two-point
functions, notably the U(1) problem., and the $\pi N$ $\sigma$ term
and the proton spin problem which involve quark loop contributions
in the three-point functions.

Instead of waiting for the advent of more powerful hardware,
there have been several suggestions
to solve it approximately with the stochastic approach. One is the
pseudo-fermion method \cite{dk86}
the other is the stochastic estimation with the Gaussian
noise~\cite{bit89}.
We shall pursue the avenue of stochastic estimation with various
noises to see if one kind of noise is better than the others.

The idea of introducing noises is hardly new in physics. Historically,
it was introduced to account for the Brownian motion with the Langevin
and Fokker-Planck equations, to compute the time-dependent correlation
functions in statistical mechanics~\cite{hh77}, for the stochastic
formulation of quantum mechanics~\cite{nel66} and quantum field
theory~\cite{pw81}.
Stochastic approach to estimating the inverse of
an $N \times N$ matrix M entails the introduction of an ensemble of
L column vectors $\eta \equiv {\eta^1,...,\eta^L}$ (each of
dimension $N \times 1$) with the properties of a white noise, i.e.
\begin{equation}\label{noise}
 \langle\eta_i\rangle = 0 ,
\hspace{0.5in}
 \langle\eta_i \eta_j\rangle = \delta_{ij},
\end{equation}
where the stochastic average $\langle \cdot \cdot \cdot \rangle$ goes
over the ensemble of the noise vectors L; e.g. $\langle \eta_i \eta_j
\rangle = \frac{1}{L} \sum_{n=1}^{L} \eta_i^n \eta_j^n$ where
$\eta_i^n$ is the i-th entry in the noise vector n.  The expectation
value of the
matrix element $M_{ij}^{-1}$ can be obtained by solving for $X_i$ in
the matrix equations $MX = \eta$ with the L noise vectors $\eta$ and
then take the ensemble average with the j-th entry of $\eta$
\begin{equation}
E[M_{ij}^{-1}] = \langle \eta_j X_i\rangle = \sum_k M_{ik}^{-1} \langle
 \eta_j \eta_k \rangle = M_{ij}^{-1}.
\end{equation}
which is the matrix element $M_{ij}^{-1}$ itself. The last step was
obtained through eq. (\ref{noise}).

The pseudofermion method~\cite{dk86} which is based on the Gaussian
distribution yields identical results as the stochastic estimation.
But the stochastic algorithm is not limited to the Gaussian noise. Any
noise which satisfies eq. (\ref{noise}) will work and one may
work better than another. To see how good a noise is, we study the
deviations from the orthonormal condition in eq. (\ref{noise}) which
is strictly true for $L \rightarrow \infty$. For
this purpose, we define the following 2 errors for measuring the
efficiency of a noise. The first one is the average absolute value
of the off-diagonal element and the second is
the deviation of the diagonal element from unity
\begin{eqnarray}
  C1&=&\frac{1}{N(N-1)}\sum_{i\neq j}\mid\langle \eta_i \eta_j \rangle
  \mid ,      \\
  C2&=&\sqrt{\frac{1}{N}\sum_i(\langle \eta_i\eta_i\rangle -1)^2} .
\end{eqnarray}
For large and independent configurations, we expect
the central limit theorem to hold, i.e.
$ C_i \cong \frac{\sigma_i}{\sqrt{L}}, i=1,2$, where  $\sigma_i$
is the standard deviation.
We checked and found this relation to hold well for the Gaussian
noise, noises with double hump distributions like
$ \eta^2 e^{-\eta^6/2}$ and $\eta^4 e^{-\eta^{10}/2}$,
and the $Z_2$ microcanonical noise $\delta(\mid\eta_x\mid -1 ) $
for the dimension $N = 500$. The results of $\sigma_1$ and $\sigma_2$
fitted from the range of $L = 10$ to 150 are given in Table 1.

\begin{table}[h]
\begin{center}
\caption{$\sigma_1$ and $\sigma_2$ as obtained from fitting the $C_1$
and $C_2$ for different noises at $N =500$ with noise configurations in
the range of $L = 10$ to 150.}
\vspace{0.5cm}
\begin{tabular}{|c||c|c|}
\hline
 noise         & $\sigma_1$       &$\sigma_2$       \\ \hline
Gaussian       & 0.78(1)          & 1.49(1)         \\ \hline
$\eta^2 e^{-\eta^4/2}$ & 0.79(1)  & 0.43(1)           \\ \hline
$\eta^4 e^{-\eta^{10}/2}$&0.80(1) & 0.28(1)           \\ \hline
$Z_2$          & 0.77(1)          & 0.00(0)          \\ \hline
\end{tabular}
\end{center}
\end{table}

We see from Table 1 that the off-diagonal deviation $\sigma_1$ is
about the same for all these noises and is close to the asymptotic
value of $\sqrt{2/\pi}$~\cite{thr92}. On the other hand,
the diagonal deviation $\sigma_2$ which depends on the deviation of
the higher moment; i.e. $\sigma_2 = \sqrt{\langle \eta^4\rangle -1}$
decreases as the distribution tends to the bi-nodal form. It is the
largest for the Gaussian case with an asymptotic value of $\sqrt{2}$
and vanishes for the $Z_2$ noise. For this reason, we suspected that
the $Z_2$ microcanonical noise may work better than the other noises
considered here~\cite{dl92}. In fact, it has been shown recently
{}~\cite{bmt93} that the variance of a inverted matrix
element due to the stochastic estimation is composed of two parts

\begin{equation}  \label{eq.variance}
Var[M_{ij}^{-1}] = \frac{1}{L} \{[M_{ij}^{-1}]^2 C_2^2 +
\sum_{k \neq j} [M_{ik}^{-1}]^2\}.
\end{equation}
Whereas the second part is independent of the kind of noise used,
the first part is proportional to the square of the
diagonal error $C_2$ only. Since $Z_2$, or $Z_N$ for that matter,
has no diagonal error, i.e. $C_2 = 0$, it produces a {\bf minimum}
variance. Other noises will have larger variances
due to the non-vanishing $C_2$. For comparison of different noises,
we consider the calculation of chiral condensate which is the trace
the inverse quark matrix M, i.e.
$\langle \overline{\Psi}\Psi\rangle = Tr M^{-1}/V$, for a quenched
$16^3 \times 24$ lattice at $\beta = 6.0$ with $\kappa = 0.148$ for the
Wilson action. First we used the conjugate gradient program to invert
a column of the matrix for a particular gauge configuration and  find
$  \sum_{k \neq 1} [M_{k1}^{-1}]^2 /[M_{11}^{-1}]^2 = 0.8$.
Assuming this ratio is true for all the
other columns based on translational invariance
and extending eq. (\ref{eq.variance}) to the
variance of the trace, we find that the standard deviation from the
$Z_2$ noise is smaller than that from the Gaussian noise by a factor
of 1.54. In other words, in order to achieve the same level of accuracy
, one would need a Gaussian noise configuration
2.4 times larger than that of the $Z_2$ noise. Since any noise with
$C_2 \neq 0$ will need more statistics than the $Z_2$ noise to
reach the same accuracy, the $Z_2$ (or $Z_N$) noise is
the optimal choice in this sense.

With the error analysis in eq. (\ref{eq.variance}), we now realize
where the stochastic inversion algorithm would apply. Since the
quark propagator has the generic fall off behavior
$e^{-m |x - y|}/(|x - y| + O(a))^n$ (a is the lattice
spacing) in the space-time separation with $n = 3$ for
short distances, quark loops that involve traces near the
diagonal, i.e $|x - y| \sim a$, will have large signals.
As long as the far off-diagonal contribution to the variance (the
second term in eq. (\ref{eq.variance})) does not overwhelm the
contribution from $[M_{ij}^{-1}]^2$ (the first term in eq.
(ref{eq.variance}), the square of the matrix element of interest, the
noise to signal ratio will be of the order $\frac{1}{\sqrt{L}}$ for the
near-diagonal traces from eq. (\ref{eq.variance}).
For the trace itself, there is an extra factor of
$\frac{1}{\sqrt{N}}$ due to the translational, gauge and rotational
symmetries. This is certainly true for the case when the quark mass
m is not very small. For $m \rightarrow 0$, it remains to be seen if
the inverse power behavior $1/(|x - y| + O(a))^n$ is steep enough to
curb the off-diagonal contribution to give a sufficiently small
variance for a reasonable L. But when one considers the case when
$|x -y| >> a$, the signal drops exponentially while the error remains
constant. Hence, the noise to signal ratio grows
exponentially and the application of the stochastic method is
invalidated under this circumstance. Therefore, one does not expect
the stochastic method to be useful for calculating general quark
propagators. But it is useful for approximating diagonal and
near diagonal traces of the inverse matrix if the inverse matrix
itself is dominated by the diagonal and near diagonal terms.

To illustrate how the $Z_2$ noise works in detail,
we employ it to invert the quark matrix on a quenched
 $16^3\times 24$ lattice at $\beta = 6.0$ for Wilson fermions
($\kappa = .148$) for a particular gauge configuration. We shall
report the results of certain diagonal and off-diagonal traces
corresponding to the scalar, the pseudoscalar, the
vector, and the axial-vector currents. The point-split currents are used
for the vector and the axial-vector cases. The accumulated averages
for these currents (summed over the spatial points on a time slice)
 for a gauge configuration are plotted in Fig. 1 against
the noise configuration L. This shows how they approach
the equilibrium up to  $L = 200$. Also plotted are the histograms for
their distributions. We see that in most cases they do tend to stabilize
for L = 200. If and when L = 200 is sufficient for the purpose of
estimating these traces with acceptable errors, this algorithm would
save the computing time by a factor of 5838 as compared to
inverting the full matrix (dimension $N = 1.18 \times 10^6$ in
this case) with the brute force approach.

   The $Z_2$ noise has bee employed to calculate quark condensate
$\langle \overline{\Psi}\Psi \rangle
= Tr M^{-1} /V$ and the topological susceptibility  with
the fermion method~\cite{sv87}, i.e.  \\
\mbox{$\chi = \frac{m^2}{V}
\langle Tr (\gamma_5 M^{-1}) Tr(\gamma_5 M^{-1}) \rangle$.} The
preliminary results~\cite{dl92} give reasonable errors which are a
combination of the errors from the stochastic estimation and the
gauge configurations. Much like the situation with the glueball
masses, we found that the two-point functions for the disconnected
quark loops are too noisy due to the exponential fall off of
the two-point function $e^{-Mt}$ as a function of t.
However, the disconnected insertion
for the three-point function is different. Since there is only one
quark loop correlated with the nucleon propagator say, the situation
is not as bad as the two-point function. Preliminary results on the
$\pi N \sigma $ term and the flavor-singlet axial charge $g_A^0$ are
encouraging~\cite{dl93} and will be reported elsewhere~\cite{dl931}.

It is worthwhile noting that the stochastic estimation is
particularly successful for the trace (denotes as $Re \overline{\Psi}
\Psi$ in Fig. 1). As we remarked before, this is due to the
translational, color and spin symmetries. As a result, the error
is proportional to $1/\sqrt{N}$ where N is the dimension of the matrix.
With $N = 1.18 \times 10^6$ in our case and  the ratio
$  \sum_{k \neq 1} [M_{k1}^{-1}]^2 /[M_{11}^{-1}]^2 = 0.8$, we
predict the error to signal ratio to be $\sqrt{1.5} \times
10^{-3}$ from eq. (\ref{eq.variance}) for L = 1. This
agrees well with the numerical calculation shown in Fig. 1.
Given this level of accuracy, it is feasible to apply the stochastic
method to the calculation of the determinant, the eigenvalues, and
the eigenvectors of the matrix M which might not be feasible with other
algorithms. It is well known that for a Hermitian matrix M, the density
of states can be written as
\begin{equation}   \label{density}
\rho(\lambda) = - \frac{1}{\pi} \lim_{\varepsilon \rightarrow 0}
Im\,Tr G(\lambda + i\varepsilon)
\end{equation}
where G is the inverse matrix of $\lambda + i\varepsilon - M$.
The determinant of the matrix M can be given as
\begin{equation}
det M = e^{\int \rho(\lambda) \ln \lambda d\lambda}
\end{equation}

Since the stochastic method is most successful in estimating the
trace of the inverse matrix M for lattice QCD as we have just remarked
and it is known that the eigenvalues of M are distributed in a
reasonably finite range, it would be worthwhile exploring the
possibility that this could be
an efficient algorithm for calculating the determinant. Looking
for the poles of the density of states in eq. (\ref{density})
has been frequently used as a way to identify the eigenvalues
{}~\cite{ww81} and the eigenvectors can be obtained from the column of
the inverse matrix
\begin{equation}
{\sl v}_i \sim \lim_{\varepsilon \rightarrow 0} Im\, G_{ik}(\lambda
+ i\varepsilon)
\end{equation}
for any column $k$~\cite{ww81}.

In conclusion, we have proposed an stochastic algorithm for
large matrix inversion with the optimal $Z_2$ noise.
We show that it is particularly efficient for estimating traces and
near diagonal traces for matrices whose inverses are dominated
by the diagonal and near diagonal terms themselves.
It is applied to calculate quark loop correlations in the vacuum and
disconnected quark loop insertions in the three-point functions in
QCD. Noting that it is most successful in estimating traces, we
shall explore the feasibility of calculating determinants, eigenvalues,
and eigenvectors in the future.

   This work was supported in part by the U.S. Department of Energy
under grant number DE-FG05-84-ER40154. We would like to thank
C. Thron for sharing his results with us prior to publication and for
stimulating discussions.

% the bibliography comes at the end

Figure Caption

\noindent
Fig. 1  The first and second columns show the accumulated averages
of the real and imaginary parts of various current loops as functions of
the noise configurations L for a gauge configuration. The third and
last columns
give the corresponding histograms for their distributions.
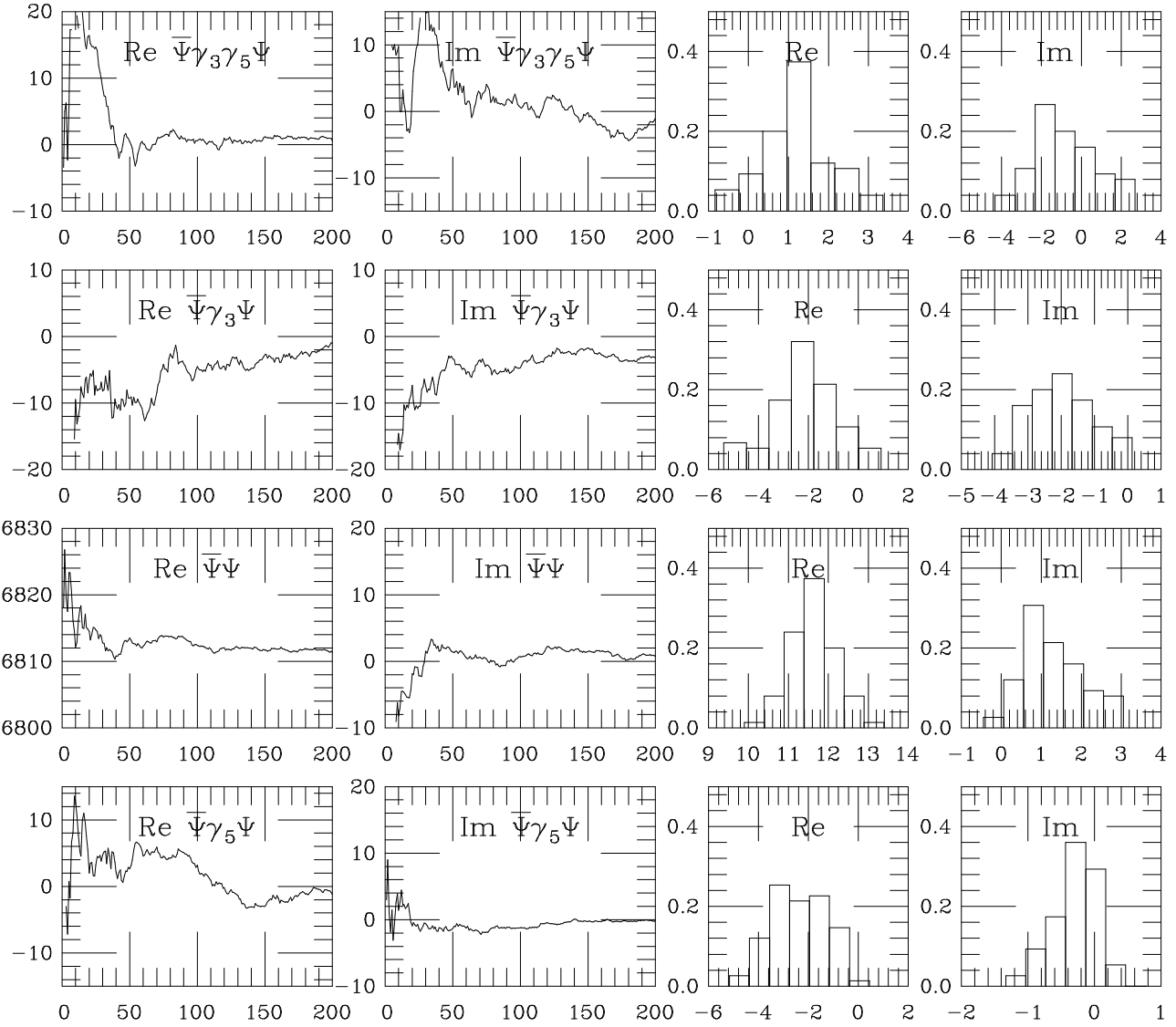
\end{document}